\documentclass[twocolumn,showpacs,preprintnumbers,amsmath,amssymb]{revtex4-1}
\usepackage{graphicx}
\usepackage{dcolumn}
\usepackage{bm}
\usepackage{color}
\begin{document}
%
%
\title{On the physical origin of fatigue phenomena}
\author{John Y. Fu}
%
%
\date{\today}
\affiliation{Department of Mechanical and Aerospace Engineering, The State University of New York, Buffalo, New York, 14260, USA}
%
%
%
\begin{abstract}
The fractional power-law material behavior has been investigated within the framework of a modified mean field theory, in which high-temperature structure precursors in a crystalline or polycrystalline material are treated as a partially ordered liquid over a wide range of temperature. By doing so, the relaxation behavior of the material can be studied via a modified Landau-Khalatnikov equation. It then becomes clear that, as the special cases of the fractional power-law behavior, seemingly quite different fatigue phenomena and physical deterioration are governed by a fundamental physical phenomenon, i.e., the competition between ordered structures and partially ordered high-temperature structure precursors during a series of atomic relaxation processes.
\end{abstract}
\maketitle
%
%
\section{Introduction}
Most material properties are often characterized by a linear relationship between the applied field and the material response. For example, the stress-strain relationship of a test specimen within the elastic range can be written as $\sigma=E\epsilon$, here $\sigma$ and $\epsilon$ represent the applied stress and the induced strain, respectively; $E$ is defined as the elastic modulus. In many cases, however, such a linear relationship could deteriorate into a fractional power-law relationship. For instance, the stress-strain relationship of the test specimen undergoing plastic deformation might be $\sigma=K_{sh}\epsilon^{n}$ (Hollomon's equation \cite{hollomon1945}), here $n$ is defined as the strain hardening exponent and $K_{sh}$ is usually called the strength index; in practice, for most metallic materials, $n$ generally varies between 0.2 and 0.5 \cite{mbm2008}. Such examples could be multiplied indefinitely. Generally speaking, the fractional power-law behavior is very often connected to physical deterioration in solid materials. In one specific field, material fatigue, this trend is particularly evident. Over the past two hundred years, material fatigue has been extensively studied and many fatigue laws have been developed. Those laws, such as the Coffin-Manson law \cite{coffin1954,manson1953}, the Basquin equation \cite{basquin1910}, the W$\mathrm{\ddot{o}}$hler equation or the W$\mathrm{\ddot{o}}$hler curve \cite{schutz1996,lalanne2002}, etc., are empirical relationships and their validity and reliability have been tested and verified by numerous experimental data. But their microscopic physical origins have not been clarified. It is interesting to notice that the Coffin-Manson exponent has shown remarkable universality in single-phase metallic materials \cite{coffinmanson1992}, which has raised a question if there is a fundamental physical phenomenon underlying apparently dissimilar fatigue phenomena in different materials. Surprisingly, the above-mentioned fatigue laws are also fractional power-law relationships, which provides a perfect opportunity to check whether there exists a universal phenomenon that governs the fractional power-law material behavior. If it does exist, then, at least, those fatigue laws should have the same origin. In this paper, we will try to give our answer to this question.
\section{Statistical interpretation of disordered structures}
It is well known that physical deterioration and fatigue phenomena in solid materials always involve a large amount of disordered microscopic structures. Therefore, a good description of those material behavior should be able to provide a statistical interpretation to explain the emergence and evolution of disordered structures in the test specimen undergoing external loading. It is perhaps worth briefly mentioning how statistical methods have been used to study the occurrence of fracture, which is slightly different from our studies. Some methods, such as fractals, scaling laws, percolation networks, etc., have been exploited to investigate fatigue and fracture phenomena \cite{disorderedmedia1990,charmet1990}. These advanced statistical methods have successfully resulted in some theoretical models that can be used to predict fracture and cracking behavior in materials; the predicted results by these models are in good agreement with experimental data to some extent. For instance, the percolation theory was used to describe the occurrence of fracture as a critical phenomenon \cite{disorderedmedia1990} and such a conceptual model was supported by some experimental results, like that reported in Ref. \cite{ciliberto1997}. The problem associated with these statistical methods is that disordered structures inside materials are neither purely stochastic in the time domain nor randomly distributed in the spatial domain over a wide range of temperature. Thus, those models may not be able to correctly describe how disordered structures alter material behavior and eventually lead to the occurrence of fracture and crack. In our opinion, any statistical method that can correctly describe that kind of material behavior must involve the consideration of the cooperative behavior or the self-organization of disordered structures.

\begin{figure}[h!]
\begin{center}
\includegraphics[width=1.0\columnwidth]{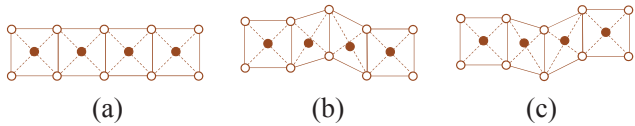}
\caption{Schematic representation of the emergence of distorted crystal lattice; (a) the normal crystal lattice without distortion; (b) the distorted crystal lattice; (c) another version of the distorted crystal lattice. Obviously, both (b) and (c) have an equal probability of occurring.}
\label{distortedlattice}
\end{center}
\end{figure}

In this study we will try another route to tackle disordered structures. Before further discussing our approach, we need to take a close look at how disordered structures are formed under thermal fluctuations. It is believed that a crystalline material has a purely ordered structure at absolute zero and a completely disordered structure near its melting point; at temperatures far below the melting point, the considered material can by no means be assumed to be perfectly ordered. As diagrammatically shown in Fig. [1], there always exists the probability that certain atoms in the normal crystal lattice, which is shown in Fig. [1a], could gain extra kinetic energy to move quasi-permanently away from their original equilibrium positions and distort crystal lattice as shown in Figs. [1b] and [1c] due to thermal fluctuations. The local disordered structures corresponding to such distorted crystal lattice are defined as high-temperature structure precursors (HTSPs). At new equilibrium positions, the distorted lattice possesses higher potential energy due to the induced local strain. Thus, HTSPs associated with such lattice distortion are often metastable: they could disappear or even ``{\it hop}" to other locations under thermal fluctuations. Since the structures shown in Figs. [1b] and [1c] have an equal probability of occurring, they can also switch from one to the other in either direction under external perturbations.

We now consider whether HTSPs could behave cooperatively. Since the formation of HTSPs is caused by thermal fluctuations, their quantity is proportional to temperature. Therefore, as temperature rises, the quantity also increases. At a certain temperature $T_{N}$, the quantity of HTSPs reaches a threshold so that they will start to interact with each other, which can be regarded as their cooperative behavior or self-organization, and then form a unique nematic phase. The driving force behind this structural transformation is the competition between energy and entropy. Let's consider a single-phase metallic material containing HTSPs. Its Gibbs free energy in the absence of external fields can be written as $G=U-TS$; clearly, the internal energy $U$ is increased due to the increment of the local strain energy generated by the distorted lattice shown in Fig. [1b] or [1c]. If temperature $T$ remains unchanged or changes slowly, the entropy $S$ must rise to reduce $G$. The simplest way to increase $S$ is that the chemical bonds between the atoms at the distorted lattice points are partially broken so that the corresponding disordered structures could gain more freedom to rotate and are oriented along local preferred directions to form a nematic phase, which leads to a decrease in the orientational entropy but an increase in the positional entropy and, eventually, results in a net increase in the total entropy. This kind of structural transformation is very common in the first order nematic-isotropic phase transition in liquid crystals \cite{degennes1995,blinov2010}. For the distorted lattice shown in Fig. [1b] or [1c], its distortion direction has an equal probability of pointing parallel or anti-parallel to a given direction (for instance, one of vertical directions shown in Fig. [1b] or [1c]). If we use $v$ to represent this given direction and approximately treat those HTSPs as molecules, then their orientational order parameter, $S_{op}$, can be written as \cite{lubensky2000}
\begin{equation}
S_{op}=\frac{1}{2}\langle3\left(v^{i}, \vec{n}\right)^{2}-1\rangle=\frac{1}{2}\langle\left(3\mathrm{cos}^{2}\theta^{i}-1\right)\rangle,
\label{orderparameter}
\end{equation}
where $\langle\ \rangle$ represents the average; $v^{i}$ is defined as the given direction of the disordered structure located at the position $i$; $\vec{n}$ is usually called the director that represents a particular direction; $\theta^{i}$ is defined as the angle between $v^{i}$ and $\vec{n}$ at the position $i$. If $0<S_{op}<1$, we can say that HTSPs cooperatively form a nematic phase. $S_{op}=1$ corresponds to an ideal case, in which all HTSPs are perfectly aligned. If temperature continues to rise, at a certain point $T=T_{NI}$, the corresponding thermal energy will be large enough to disturb HTSPs, which forces HTSPs to be randomly oriented and $S_{op}=0$. Therefore, when $T>T_{NI}$, all HTSPs behave like a normal liquid, which has an isotropic phase. In this paper, we only consider the case, in which $T_{N}<T<T_{NI}$.

In the above discussion, we use an orientational order parameter to represent, in the statistical sense, the cooperative behavior of HTSPs. Before giving further discussion, we have to briefly introduce the concept of the order parameter. In 1937, Landau introduced a specific thermodynamic variable, i.e., the order parameter, to describe his idea of broken symmetry in continuous phase transitions \cite{landau1937}. In the absence of the external field and in the vicinity of the critical point, $T_{c}$, the Landau free energy $F$ of a unit-volume material can be written as follows \cite{landau1937}
\begin{equation}
F=F_{0}+\frac{1}{2}aP^{2}+\frac{1}{4}bP^{4}+\cdots,
\label{landau1937}
\end{equation}
where $P$ is defined as the order parameter and $F_{0}$ is the free energy that is independent of $P$; both $a$ and $b$ are coefficients; Within the framework of the Landau theory, $P$ has the following values: $P=0$ if $T>T_{c}$, which corresponds to one phase with higher symmetry, and $P=\pm \ constant$ if $T<T_{c}$, which corresponds to another phase with lower symmetry. If temperature $T$ continually changes, we can say that the material undergoes a continuous phase transition when $T$ passes through the critical point. In this study, we generalize Landau's concept of the order parameter to represent the collective response of a material when disturbed by external fields. If a uniform stress $\sigma$ is applied to a unit-volume metallic material, the collective response, or strain $\epsilon$, of the material can be defined as its order parameter. We now consider the behavior of the above-defined nematic phase under the perturbation of $\sigma$. Since HTSPs are usually metastable, $\sigma$, just like thermal fluctuations, will disturb them. According to Le Chatelier's principle, the nematic phase as the ensemble of HTSPs would then undergo a specific structural change, in which HTSPs would tend to huddle together and grow up, to counteract any imposed deformation by $\sigma$ \cite{landau1980}. Therefore, the effective macroscopic stress inside the material will be $\sigma_{eff}=\sigma-\tilde{\alpha}\epsilon_{dis}$; here the negative sign is due to Le Chatelier's principle, $\epsilon_{dis}$ is the effective strain induced by HTSPs in the nematic phase, and $\tilde{\alpha}$ is the coefficient related to the nematic phase. For convenience, we can further write $\sigma_{eff}=\sigma-\tilde{\alpha}\epsilon_{dis}=(1-k)\sigma$ and define $k$ as
\begin{equation}
k=\frac{k_{0}S_{op}(T-T_{N})}{T_{N}} \ \ \ \mbox{$(T_{N}<T<T_{NI})$},
\label{niphasetransition}
\end{equation}
where both $k$ and $k_{0}$ are dimensionless coefficients. Then the Gibbs free energy of that unit-volume metallic material can be written as
\begin{equation}
G=U-TS-(1-k)\sigma\epsilon=U-TS-(1-k)E\epsilon^{2},
\label{modifiedgibbs}
\end{equation}
where one can see that $k$ actually represents, in the statistical sense, the fraction of the potential energy, which is generated by the cooperative movement of HTSPs. Now we can briefly summarize our approach: both the crystalline phase and the nematic phase as the ensemble of HTSPs in a test specimen are integrated together by using a modified mean field theory (MMFT), in which the former corresponds to the collective response of the crystalline phase to external stimuli and the latter the extremely slow counteractive and fluctuating response of the nematic phase of HTSPs in the specimen.

If we assume that $\sigma$ is a function of time, then, as time goes, $k$ will increase due to Le Chatelier's principle. Thus the modulated strain $\sqrt{1-k}\epsilon$ is also a time-varying value. Clearly, for a perfect single-phase crystalline material, $k\approx0$ at the beginning when the external force is applied. At this moment, it is safe to write down the relationship $\sigma=E\epsilon$ if the value of $\sigma$ is small. However, $k$ will continue to increase under the perturbation of the time-varying external force. In order to determine how the evolution of $k$ could eventually alter the linear stress-strain relationship, we need to exploit the Landau-Khalatnikov equation to investigate the relaxation behavior of the modulated strain.
\section{Fractional power-law material behavior in characteristic relaxations}
In 1954, Landau and Khalatnikov formulated an important equation, which is written below \cite{landau1954},
\begin{equation}
\gamma\frac{dP}{dt}=-\frac{\partial F}{\partial P},
\label{lk1954}
\end{equation}
where $\gamma$ is a kinetic coefficient and is considered to be independent of temperature \cite{blinov2010,blinc1974}; $t$ represents the time variable; the definitions of both $F$ and $P$ have been given in Eq. (\ref{landau1937}). This equation is usually called the Landau-Khalatnikov equation that was originally developed to describe the critical slowing down of the fluctuation of the order parameter on approaching the critical point \cite{blinc1974}. But it has a more profound physical significance; it can be interpreted as another version of the law of conservation of energy, i.e., the kinetic energy associated with the fluctuation of the order parameter and dissipated during the corresponding relaxation process is equal to the decrease in the Landau free energy. Therefore, this equation can be exploited to analyze the relaxation behavior of the order parameter in general cases. In this study, we replace Landau's concept of the order parameter (the spontaneous strain) by a generalized order parameter (the induced strain). The uniqueness of our approach is that both the above-mentioned collective and slow counteractive responses can be integrated together in the mathematical expression of the Gibbs free energy to represent the competition between the crystalline phase and the nematic phase in the test specimen. Therefore, we can investigate this competition or, in other words, how the time-varying $k$ could eventually alter the linear stress-strain relationship via relaxation processes.

We now need to modify the Landau-Khalatnikov equation for our studies. Notice that $\frac{1}{2}aP^{2}$ does positive contribution to the Landau free energy given in Eq. (\ref{landau1937}) but $(1-k)E\epsilon^{2}$ does negative contribution to the Gibbs free energy given in Eq. (\ref{modifiedgibbs}), so we have to modify the Landau-Khalatnikov equation as follows.
\begin{equation}
\gamma\frac{d\epsilon}{dt}=\frac{\partial G}{\partial\epsilon}.
\label{modifiedlk2012a}
\end{equation}
Substitute $G$ given in Eq. (\ref{modifiedgibbs}) into the above equation, we get the following result.
\begin{equation}
\gamma\frac{d\epsilon}{dt}=\frac{\partial G}{\partial\epsilon}=-2(1-k)E\epsilon.
\label{modifiedlk2012b}
\end{equation}
From this equation, we can see that the SI unit of $\gamma$ is $\mathrm{N\cdot s/m^{2}}$, which is similar to that of dynamic viscosity (in fact, it has been pointed out that $\gamma$ is viscosity for liquid crystals \cite{blinov2010}). Solve this equation, we have,
\begin{equation}
\epsilon=\epsilon_{s}\mathrm{exp}\left[-\frac{2(1-k)E}{\gamma}t\right],
\label{modifiedlk2012c}
\end{equation}
where $\epsilon_{s}=\epsilon|_{t=0}$, which can be interpreted as the induced strain under static deformation. For simplicity, we call $\epsilon_{s}$ the static strain. Eq. (\ref{modifiedlk2012c}) represents the relaxation behavior of the induced strain after an instantaneous external perturbation. Since the relaxation process considered here is ultimately connected to the motion of atoms in crystalline lattices or partially ordered structures, we would introduce an atomic relaxation time, $\tau$, to characterize the relaxation behavior at the atomic level. Now we can approximately estimate the magnitude of the parameters given in Eq. (\ref{modifiedlk2012c}). For a fully relaxed process, we could regard $t$ as $t\approx\tau$; since $\tau\sim\frac{1}{f_{D}}$, where $f_{D}$ is defined as the Debye frequency that is about $f_{D}\sim10^{13}\mathrm{Hz}$ \cite{am1976}, so $t\sim10^{-13}\mathrm{s}$; for most metallic materials, $E$ ranges from $10^{9}\mathrm{N/m^{2}}$ to $10^{11}\mathrm{N/m^{2}}$; $k$ ranges from 0 to 0.5 (we will discuss this later); since $\gamma$ represents the kinetic behavior at the atomic level, it is reasonable to assume that $\gamma\gg1$. Actually, $\gamma$ could be regarded as the viscosity, which describes the viscous behavior of the nematic phase of HTSPs; therefore, it must be very large. Thus we can estimate that $|-\frac{2(1-k)E}{\gamma}t|\ll1$. By using Eq. (\ref{maclaurin1}) in Appendix, we linearize Eq. (\ref{modifiedlk2012c}) as follows
\begin{equation}
\epsilon\approx\epsilon_{s}\left[1+\frac{2(k-1)E}{\gamma}t\right].
\label{modifiedlk2012d}
\end{equation}

The above equation only represents a single relaxation after an instantaneous external perturbation; if we want to use it to investigate fatigue phenomena, the special cases of the fractional power-law material behavior, we must find a series of such atomic relaxations to form a continuous and progressive relaxation process in the test specimen when it is subject to cyclic loading. For this purpose, we consider three loading signals that are shown schematically in Fig. [2]. The first signal S-I, one period or cycle of it is shown in Fig. [2a], is a one-direction loading signal; there is no reversal in such a signal so that its cycle number is usually used to demonstrate fatigue phenomena. If we define $T_{p}$ as the period of S-I, in the first half of $T_{p}$, a constant load is applied; no load in the second half. If $T_{p}\gg\tau$, the atomic relaxation will fully occur once at $\frac{T_{p}}{2}$. The second loading signal S-II, one period or cycle of it is shown in Fig. [2b], is a fully reversed loading signal; there are two reversals in one cycle of S-II and its number of reversals is usually used to demonstrate fatigue phenomena. If we also define $T_{p}$ as the period of S-II, in the first half of $T_{p}$, a constant load is applied; in the second half, the same load but in the opposite direction is applied. If $T_{p}\gg\tau$, the atomic relaxation will fully occur twice at both $\frac{T_{p}}{2}$ and $T_{p}$, respectively. The third loading signal S-III, shown in Fig. [2c], is a constant loading signal; there is no reversal in S-III and the atomic relaxation does not occur during the time in which S-III is applied. For simplicity, we will mainly choose S-I with $N$ cycles as the applied loading signal in our studies. Since both S-I and S-II are commonly used in fatigue studies in practice, we will later convert our derived formulas to the ones based on S-II via $\epsilon\rightarrow\epsilon/2$ and $N\rightarrow2N$. In this way, we will be able to present the fatigue relationships connected to both S-I and S-II, respectively. Here we have to emphasize that the fatigue phenomena connected to S-I might be different from the ones connected to S-II. The reason for this can be explained in such a way that the value of $k$ increases more quickly under the S-II loading test than that under the S-I loading test.

Because the applied loading signal S-I has $N$ cycles, $t$ given in Eq. (\ref{modifiedlk2012d}) should be re-counted as $t=N\tau$. Substitute it into Eq. (\ref{modifiedlk2012d}), we have the following relationship.
\begin{equation}
\epsilon\approx\epsilon_{s}\left[1+(k-1)\rho\tau N\right],
\label{modifiedlk2012e}
\end{equation}
where $\rho=\frac{2E}{\gamma}$. This equation represents the aforementioned continuous and progressive relaxation process occurring in the test specimen when it undergoes a cyclic loading test. We now try to use it to derive different fatigue relationships. If assuming $|\rho\tau N|\ll1$, then, according to Eq. (\ref{binomial}) in Appendix, we can further simplify Eq. (\ref{modifiedlk2012e}) and get a new result as follows
\begin{eqnarray}
\epsilon & \approx & \epsilon_{s}(1+\rho\tau N)^{k-1}=\epsilon_{s}\left(\frac{1}{N}+\rho\tau\right)^{k-1}N^{k-1} \nonumber \\
& = & \epsilon_{sm}N^{-(1-k)}=\epsilon_{sm}N^{-\beta},
\label{coffinmanson1}
\end{eqnarray}
here $\epsilon_{sm}=\epsilon_{s}\left(\frac{1}{N}+\rho\tau\right)^{k-1}$. Making the following conversion, $\epsilon\rightarrow\epsilon/2$ and $N\rightarrow2N$, we get the same relationship but based on S-II, which is given below.
\begin{equation}
\frac{\epsilon}{2}\approx\epsilon_{sm}(2N)^{-(1-k)}=\epsilon_{sm}(2N)^{-\beta}.
\label{coffinmanson2}
\end{equation}
This equation describes a general relationship between $\epsilon$ and $N$; here $N$ may not be the fatigue life $N_{f}$. When $N<N_{f}$, both $\epsilon_{sm}$ and $\beta$ are functions of $N$. Notice that we did not mention the amplitudes of the loading signals defined in Fig. [2]. Therefore, $\epsilon$ and its corresponding applied stress are chosen arbitrarily. Now, if we choose an applied stress that is large enough for plastic deformation to occur in the test specimen, a fixed fatigue life $N_{f}$ can be found to correspond to the induced plastic strain $\epsilon_{p}$. Then we can get the following equation
\begin{equation}
\frac{\epsilon_{p}}{2}\approx\epsilon_{f}(2N_{f})^{-(1-k)}=\epsilon_{f}(2N_{f})^{-\beta},
\label{coffinmanson3}
\end{equation}
where $\epsilon_{f}=\epsilon_{sm}|_{N=N_{f}}$, which is defined as the ductility coefficient. This equation is the mathematical expression of the Coffin-Manson law, which was proposed independently by L. F. Coffin in 1954 \cite{coffin1954} and S. S. Manson in 1953 \cite{manson1953} to describe an empirical relationship between the induced plastic deformation and the number of reversals to failure. $\beta=1-k$ given in the above equation is defined as the Coffin-Manson exponent. It has been observed that this exponent possesses a remarkable universality; $\beta\sim0.5$ has been found in single-phased metallic materials whatever crystalline or polycrystalline structures they have \cite{coffinmanson1992,lcf1988}. It is easy to see that the Coffin-Manson law is only the special case (when $N=N_{f}$ and $k\sim0.5$) of Eq. (\ref{coffinmanson2}).

Unfortunately, at this moment, there is no convincing theory that can be used to explain the above-mentioned universal law observed in the Coffin-Manson exponent \cite{coffinmanson1992}. We now try to interpret the physical meaning of this exponent within the framework of the MMFT. As we have already discussed, the effective macroscopic stress inside the test specimen is $\sigma_{eff}=\sigma-\tilde{\alpha}\epsilon_{dis}=(1-k)\sigma$. It is well known that there is an essential prerequisite for the rationality and validity of any thermodynamic model, i.e, the specimen should be a continuum at any time when it is subject to an external perturbation. In thermodynamics, material properties are studied by using continuous functions so that the considered material must be treated as a continuum \cite{lubensky2000}. Now we need to take a close look at the meaning of $\beta=1-k\sim0.5$ or $k\sim0.5$. Under the framework of the MMFT, for a single phase crystalline material, the value of $k$ characterizes the counteractive response of the nematic phase as the ensemble of HTSPs to external perturbations; whereas that of $1-k$ represents the collective response of the crystalline phase to external perturbations. Since the chemical bonds between certain atoms of HTSPs are assumed to be partially broken, they cannot be considered the integral part of the crystalline lattice of the test specimen anymore. Therefore, $k$ represents, in the statistical sense, the fraction of the total atoms that are not the integral part of the crystalline lattice in the test specimen. In other words, $k=0.5$ represents a critical point; below this point ($k<0.5$ or $\beta>0.5$), the specimen is still a continuum; above this point ($k>0.5$ or $\beta<0.5$), the specimen is not a continuum anymore. Thus, from the viewpoint of thermodynamics, the Coffin-Manson law is only valid when $k<0.5$. It might also be interesting to interpret the meaning of $\beta\sim0.5$ from the viewpoint of deformation and fracture of materials. Under a small external perturbation, the crystalline lattice will undergo harmonic vibration, which corresponds to the elastic deformation, and HTSPs will ``{\it flow}", which corresponds to the plastic deformation, in the test specimen. If the perturbation is large enough, both the crystalline lattice and HTSPs will undergo plastic deformation simultaneously in the test specimen. When $k<0.5$ or $\beta>0.5$, the induced elastic deformation dominates; thus we can say that the global deformation in the test specimen, in the statistical sense, is elastic. Of course, the local plastic deformation due to the movement of HTSPs could also be induced in the test specimen even when $k$ is very small; but this cannot alter the fact that the global elastic deformation dominates under such circumstances. When $k>0.5$ or $\beta<0.5$, however, the induced plastic deformation dominates; thus we can say that the global deformation in the test specimen, in the statistical sense, is plastic. Therefore, when $\beta=1-k\rightarrow0.5$, the global plastic deformation will occur and the fracture crack will emerge and start to grow in the test specimen. In this sense, fatigue can be regarded as the precursor to fracture. This also explains why the global plastic deformation can be observed in the Coffin-Manson law when $\beta=1-k\rightarrow0.5$ in single phase crystalline materials. In practice, since there always exist defects in real metallic materials and/or alloys and these {\it built-in} disordered structures could lower the value of $k$, $\beta$ usually ranges from 0.5 to 0.7 \cite{mbm2008}. Clearly, the Coffin-Manson law is one of the fractional power-law relationships, in which $k\neq0$ and $\beta=1-k<1$.

The fatigue phenomena described by the Coffin-Manson law are usually defined as low-cycle fatigue (LCF), in which the amplitude of loading signals is large and the number of cycles to failure is small. There is another type of fatigue, high-cycle fatigue (HCF), in which the amplitude of loading signals is very small and the number of cycles to failure is very large. We now try to derive one of important HCF laws, the Basquin equation, which was proposed by O. H. Basquin in 1910 \cite{basquin1910} to describe an empirical relationship between the elastic strain amplitude and a large number of reversals to failure. Since the number of cycles to failure in HCF is usually more than $10^{4}$ \cite{lalanne2002}, we define the cycle number used in the following derivation as $N_{h}$ so that it can be distinguished from $N$ used previously. Similarly, we let $t=N_{h}\tau$ and then substitute it into Eq. (\ref{modifiedlk2012d}). We have the following equation
\begin{eqnarray}
\epsilon & \approx & \epsilon_{s}\left[1+(k-1)\rho\tau N_{h}\right] \nonumber \\
& = & \epsilon_{s}\left[1+\eta(k-1)\rho\tau\frac{N_{h}}{\eta}\right],
\label{basquin1}
\end{eqnarray}
here $\eta$ is a coefficient, which is less than 1; we will explain why we need this coefficient later. The above equation is the corresponding continuous and progressive relaxation process occurring in the test specimen when it undergoes a HCF test. If assuming $|\rho\tau N_{h}|\ll1$, then, according to Eq. (\ref{binomial}) in Appendix, we can further simplify the above equation and get a new formula as follows
\begin{eqnarray}
\epsilon & \approx & \epsilon_{s}\left(1+\rho\tau\frac{N_{h}}{\eta}\right)^{\eta(k-1)} \nonumber \\
& = & \epsilon_{s}\left(\frac{1}{N_{h}}+\frac{\rho\tau}{\eta}\right)^{\eta(k-1)}N_{h}^{\eta(k-1)} \nonumber \\
& = & \epsilon_{sn}N_{h}^{-\eta(1-k)} \nonumber \\
& = & \frac{\sigma_{sn}}{E}N_{h}^{-\eta(1-k)}=\frac{\sigma_{sn}}{E}N_{h}^{-b},
\label{basquin2}
\end{eqnarray}
where $\epsilon_{sn}=\epsilon_{s}\left(\frac{1}{N_{h}}+\frac{\rho\tau}{\eta}\right)^{\eta(k-1)}$; $b=\eta(1-k)$ is an exponent. Since the deformation involved in HCF is often believed to be mainly elastic, $\epsilon_{sn}$ given in the above equation is usually replaced by $\frac{\sigma_{sn}}{E}$. Making the following conversion, $\epsilon\rightarrow\epsilon/2$ and $N_{h}\rightarrow2N_{h}$, we get the same formula but based on S-II, which is written as follows
\begin{equation}
\frac{\epsilon}{2}\approx\frac{\sigma_{sn}}{E}(2N_{h})^{-\eta(1-k)}=\frac{\sigma_{sn}}{E}(2N_{h})^{-b}.
\label{basquin3}
\end{equation}
This equation describes a general relationship between $\epsilon$ and $N_{h}$; similarly, $N_{h}$ may not be the fatigue life $N_{f}$ and both $\sigma_{sn}$ and $b$ are functions of $N_{h}$ when $N_{h}<N_{f}$. Now, if we choose an applied stress that can only induce elastic deformation in the test specimen at the initial stage, a long fatigue life $N_{f}$ can be found to correspond to the induced strain $\epsilon_{e}$. Then we can write the following equation
\begin{equation}
\frac{\epsilon_{e}}{2}\approx\frac{\sigma_{f}}{E}(2N_{f})^{-\eta(1-k)}=\frac{\sigma_{f}}{E}(2N_{f})^{-b},
\label{basquin4}
\end{equation}
where $\sigma_{f}=\sigma_{sn}|_{N_{h}=N_{f}}$, which is defined as the fatigue strength coefficient. This equation is the mathematical expression of the Basquin equation and $b=\eta(1-k)$ is defined as the Basquin exponent for convenience. In practice, $b$ ranges from 0.125 to 0.2 \cite{mbm2008}. It is clear that the Basquin equation is only the special case (when $N_{h}=N_{f}$ and $k\sim0.5$) of Eq. (\ref{basquin3}). It is also clear that the Basquin equation is one of
the fractional power-law relationships.

\begin{figure}[h!]
\begin{center}
\includegraphics[width=1.0\columnwidth]{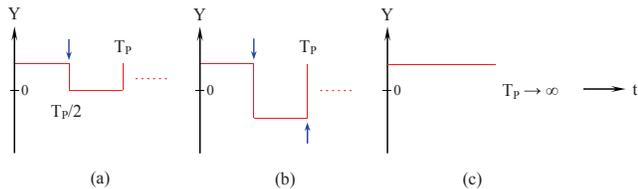}
\caption{Schematic representation of loading signals; (a) S-I: the characteristic relaxation occurs only once within one $\mathrm{T_{p}}$; (b) S-II: the relaxation occurs twice within one $\mathrm{T_{p}}$; (c) S-III: no relaxation occurs; here $\mathrm{T_{p}}$ is the period of loading signals.}
\label{loadingsignals}
\end{center}
\end{figure}

From what have been derived above, it is easy to see that there is no fundamental difference between the Coffin-Manson law and the Basquin equation. Both of them arise from the competition between the crystalline phase and the nematic phase as the ensemble of HTSPs during the aforementioned continuous and progressive relaxation process in the test specimen; when approaching the numbers of reversals to failure in both cases, $k$ is going to reach its limit, i.e., $k\rightarrow0.5$. Only $k$ varies differently in both cases; it increases quickly in the case of the Coffin-Manson law but slowly in that of the Basquin equation. Such dynamic behavior of $k$ might cause problems in the latter case. If we assume that $b=1-k$; since $k$ may vary slowly in a HCF test, there always exists the possibility that when $b=1-k$ is still around 1 ($k\sim0$) whereas $N_{h}$ has become very large but has not reached $N_{f}$, which will result in $\frac{\epsilon_{e}}{2}\rightarrow0$. In order to prevent such a problem from occurring, we have to introduce a coefficient $\eta$ into the Basquin equation to reduce the value of $b$. Obviously, for $b=\eta(1-k)$, $\eta$ must be less than 1. But the choice of the value of $\eta$ is quite arbitrary, which can also cause problems for HCF studies in practice. Probably, this is one of the reasons that the Basquin equation, even widely used, is not so accurate as the W$\mathrm{\ddot{o}}$hler equation or the W$\mathrm{\ddot{o}}$hler curve \cite{lalanne2002}, another important HCF law proposed by A. W$\mathrm{\ddot{o}}$hler in 1870 \cite{schutz1996,lalanne2002}.

We now try to derive the W$\mathrm{\ddot{o}}$hler equation. Assuming that $|(k-1)\rho\tau N_{h}|\ll1$, then, using Eq. (\ref{maclaurin2}) in Appendix, we can modify Eq. (\ref{basquin1}) and then get a new equation as follows
\begin{eqnarray}
\epsilon & \approx & \epsilon_{s}-\epsilon_{s}\mathrm{ln}\left[1-(k-1)\rho\tau N_{h}\right] \nonumber \\
& = & \epsilon_{s}\left\{1-\mathrm{ln}\left[\frac{1}{N_{h}}+(1-k)\rho\tau\right]\right\}-\epsilon_{s}\mathrm{ln}(N_{h}) \nonumber \\
& = & \epsilon_{s}\psi-\epsilon_{s}\frac{\mathrm{log_{10}}(N_{h})}{\mathrm{log_{10}}(e)} \nonumber \\
& \approx & \epsilon_{s}\psi-2.3\epsilon_{s}\mathrm{log_{10}}(N_{h}),
\label{Wohler1}
\end{eqnarray}
where $\psi=\left\{1-\mathrm{ln}\left[\frac{1}{N_{h}}+(1-k)\rho\tau\right]\right\}$. Since the above equation represents a HCF relationship, the deformation involved in this equation is usually defined as the elastic one. Thus we can multiply both sides of the above equation by $E$ and get the following result
\begin{equation}
\sigma_{w}\approx\alpha_{w}-\beta_{w}\mathrm{log_{10}}(N_{h}),
\label{Wohler2}
\end{equation}
where $\sigma_{w}=\epsilon E$, $\alpha_{w}=\epsilon_{s}\psi E$, and $\beta_{w}=2.3\epsilon_{s}E$. Similarly, this equation describes a general logarithmic relationship between $\sigma_{w}$ and $N_{h}$; here $N_{h}$ may not be the fatigue life $N_{f}$. Now, if we choose an applied stress $\sigma_{w}$ that can only induce elastic deformation in the test specimen at the initial stage, a long fatigue life $N_{f}$ can be found to correspond to $\sigma_{w}$. Then we can get the following result
\begin{equation}
\sigma_{w}\approx\alpha_{f}-\beta_{w}\mathrm{log_{10}}(N_{f}),
\label{Wohler3}
\end{equation}
where $\alpha_{f}=\alpha_{w}|_{N_{h}=N_{f}}$. This equation is the mathematical expression of the W$\mathrm{\ddot{o}}$hler equation \cite{schutz1996,lalanne2002}. One can see that the W$\mathrm{\ddot{o}}$hler equation is actually the special case (when $N_{h}=N_{f}$ and $k\sim0.5$) of Eq. (\ref{Wohler2}).

In the Basquin and the W$\mathrm{\ddot{o}}$hler equations, both $k\sim0.5$ and $N_{h}=N_{f}$ (for perfect single-phase metallic materials) demonstrate that the co-existence of the induced elastic deformation and the induced plastic deformation in the test specimen has reached the balance point, which contradicts the widely used approximation method that the induced deformation in both cases is treated as the elastic one so that the stress variables instead of the strain variables are used in both equations. In practice, such a treatment is particularly straightforward and convenient; but, in theory, it may cause some problems since the induced deformation is by no means purely elastic under those circumstances.

It might also be worth considering a special case, in which the loading signal is S-III shown in Fig. [2c]. There is no reversal in S-III and the atomic relaxation does not occur during the time in which S-III is applied. If we assume that the amplitude of S-III is large enough to induce plastic deformation in the test specimen, the value of $k$ could also reach 0.5 under static or quasi-static deformation. Since the relaxation does not occur under such deformation, without loss of generality, we let $t=\tau$ and substitute it into Eq. (\ref{modifiedlk2012c}); the result is written as follows.
\begin{equation}
\epsilon=\epsilon_{s}\mathrm{exp}\left[-2(1-k)E\frac{\tau}{\gamma}\right]=\epsilon_{s}\mathrm{exp}\left[-(1-k)\rho\tau\right],
\label{strainhardening1}
\end{equation}
here $\rho\tau$ is dimensionless; $\epsilon$ represents the induced final plastic strain and $\epsilon_{s}$ can be regarded as the initial static strain. As we have already discussed, we know that $|-(1-k)\rho\tau|\ll1$. By taking advantage of Eq. (\ref{maclaurin1}) in Appendix, we can simplify the above equation and get the following formula
\begin{equation}
\epsilon\approx\epsilon_{s}-(1-k)\rho\tau\epsilon_{s},
\label{strainhardening2}
\end{equation}
or
\begin{equation}
\epsilon_{s}\approx\epsilon+(1-k)\rho\tau\epsilon_{s}.
\label{strainhardening3}
\end{equation}
Since $\rho\tau\ll1$, by using Eq. (\ref{binomial}) in Appendix, we can further simplify the above equation and get a new result as follows
\begin{equation}
\epsilon_{s}+1\approx\epsilon+\left(\frac{1}{\epsilon_{s}}+\rho\tau\right)^{1-k}\epsilon_{s}^{1-k}.
\label{strainhardening4}
\end{equation}
This equation is the fundamental equation that can be interpreted as follows. During the time, in which S-III is applied, the value of $k$ increases as time goes. The above equation demonstrates the relationship between the initial static strain $\epsilon_{s}$ and the final plastic strain $\epsilon$; for a perfect single phase crystalline material, $k=0$ at the initial stage and $k\rightarrow0.5$ at the final stage. Now we try to alter the mathematical expression of the above equation. Simply multiplying both sides of the above equation by an effective modulus, $\tilde{E}$, we get
\begin{equation}
\sigma\approx\sigma_{y}+K_{sh}\epsilon_{s}^{1-k}=\sigma_{y}+K_{sh}\epsilon_{s}^{n},
\label{strainhardening5}
\end{equation}
where $\sigma=(\epsilon_{s}+1)\tilde{E}$, $\sigma_{y}=\epsilon\tilde{E}$, and $n=1-k$; $K_{sh}=\left(\frac{1}{\epsilon_{s}}+\rho\tau\right)^{1-k}\tilde{E}$. This equation is the mathematical expression of the Ludwik's equation, which was proposed by P. Ludwik in 1909 \cite{ludwik1909}. Here we need to give extra explanation of this equation. In practice, the time interval between the initial stage ($k=0$) and the final stage ($k\rightarrow0.5$) is short since S-III is large, therefore, we have the following approximate definitions: $\sigma$ is the stress, $\sigma_{y}$ represents the yield stress, and $\epsilon_{s}$ is regarded as the plastic strain; $n=1-k$ is usually called the strain hardening exponent. If we re-write Eq. (\ref{strainhardening4}) as follows,
\begin{equation}
\epsilon_{s}-\epsilon+1=\tilde{\epsilon}\approx\left(\frac{1}{\epsilon_{s}}+\rho\tau\right)^{1-k}\epsilon_{s}^{1-k},
\label{hollomon1}
\end{equation}
here $\tilde{\epsilon}$ approximately represents the induced plastic strain at any given time between the initial stage and the final stage. Similarly, multiplying both sides of the above equation by $\tilde{E}$, we have
\begin{equation}
\tilde{\sigma}\approx K_{sh}\epsilon_{s}^{1-k}=K_{sh}\epsilon_{s}^{n},
\label{hollomon2}
\end{equation}
here $\tilde{\sigma}=\tilde{\epsilon}\tilde{E}$. This equation is the mathematical expression of the Hollomon's equation, which was proposed by J. H. Hollomon in 1945 \cite{hollomon1945}. In practice, for most metallic materials and/or alloys, $n$ generally varies between 0.2 and 0.5 \cite{mbm2008}. The fractional power-law relationships between the applied stress and the induced strain described by both the Ludwik's and the Hollomon's equations are usually defined as the strain hardening effect. The classical explanation of this effect is that the induced plastic deformation would cause the generation of additional dislocations in the test specimen; the more dislocations, the more they would likely become pinned together, which reduces the mobility of dislocations and prevents further deformation from occurring \cite{mbm2008}. Obviously, this effect can also be explained by the model proposed here, i.e., the material hardening or strengthening is due to the fact that the aforementioned disordered structures would tend to huddle together and grow up, due to Le Chatelier's principle, to counteract any imposed deformation by the applied stress.

It is interesting to note that the strain hardening exponent $n$ ranges from 0.2 to 0.5, which corresponds to the value of $k$ varying from 0.5 to 0.8. As we have discussed, $k<0.5$ is the prerequisite for the validity of any thermodynamic model. Why is $k$ greater than 0.5 in both the Ludwik's and the Hollomon's equations? We try to explain this as follows. In order to derive these two equations, we multiply both sides of Eq. (\ref{strainhardening4}) by $\tilde{E}$. In doing so, we actually assume that there are linear relationships between $\sigma$ and $(\epsilon_{s}+k)$, between $\sigma_{y}$ and $\epsilon$, etc. Obviously, this simple approach cannot result in perfectly correct mathematical expressions since the deformation involved here is plastic. This might be one of the reasons that the measured $n$ could be less than 0.5 in practice. Theoretically, compared with both the Ludwik's and the Hollomon's equations, Eq. (\ref{strainhardening4}) may be a better model for the mathematical description of the strain hardening effect.
\section{Concluding Remarks}
In this study, the origin of the fractional power-law material behavior has been investigated. We generalize Landau's concept of the order parameter to represent the collective response of the crystalline phase of a test specimen when disturbed by external fields and treat its HTSPs as a nematic phase or a partially ordered liquid over a wide range of temperature. Both responses of the crystalline and the nematic phases to external fields are then integrated together within the framework of the MMFT. In doing so, the relaxation behavior of the generalized order parameter of the specimen can be studied via a modified Landau-Khalatnikov equation. In view of what have been derived and discussed above, one can see that the fractional power-law material behavior is closely related to the nematic phase of HTSPs; furthermore, we can conclude that, as the special cases of the fractional power-law material behavior, different fatigue phenomena are actually governed by the competition between the crystalline phase and the nematic phase as the ensemble of HTSPs during a series of atomic relaxations in the test specimen.
\begin{acknowledgements}
The research presented here was sponsored by the State University of New York at Buffalo.
\end{acknowledgements}
%
%
%
%
%
%
%
\appendix
\section{Series expansions of exponential and natural logarithm functions}
For an exponential function $e^{x}$, its Maclaurin series expansion can be written as
\begin{eqnarray}
e^{x} & = & \sum_{n=0}^{\infty}\frac{x^{n}}{n!} \nonumber \\
& = & 1+x+\frac{x^{2}}{2!}+\frac{x^{3}}{3!}+\cdots. \ \ \ (\mathrm{for\ all\ x}) \nonumber
\end{eqnarray}

If $|x|\ll1$, the above equation can be simplified as follows,
\[
e^{x}\approx1+x. \qquad \tag{A1}\label{maclaurin1}
\]

For a natural logarithm function $\mathrm{ln}\ y=\mathrm{ln}(1-x)$, its Maclaurin series expansion can be written as
\begin{eqnarray}
\mathrm{ln}\ y & = & -\sum_{n=1}^{\infty}\frac{x^{n}}{n} \nonumber \\
& = & -x-\frac{x^{2}}{2}-\frac{x^{3}}{3}-\cdots. \ \ (-1\leq x<1) \nonumber
\end{eqnarray}

Similarly, if $|x|\ll1$, the above equation can be simplified as follows,
\[
\mathrm{ln}\ (1-x)\approx-x. \qquad \tag{A2}\label{maclaurin2}
\]

\section{Binomial series expansion}
For a function $f(x)=(1+x)^{\xi}$, its binomial series expansion can be written as
\begin{eqnarray}
(1+x)^{\xi} & = & 1+\xi x+\frac{\xi(\xi-1)}{2!}x^{2}+\frac{\xi(\xi-1)(\xi-2)}{3!}x^{3}+ \nonumber \\
& & \cdots+\frac{\xi(\xi-1)\cdots(\xi-n+1)}{n!}x^{n}+\cdots. \nonumber
\end{eqnarray}

For simplicity, we here only consider the case that $\xi$ is a real number. If $|x|<1$, this expansion converges absolutely for any number $\xi$. If $|x|\ll1$, we can get the following approximation formula,
\[
(1+x)^{\xi}\approx1+\xi x. \qquad \tag{B1}\label{binomial}
\]
\end{document}